\documentclass[sigconf,anonymous=false]{acmart}

\AtBeginDocument{%
  \providecommand\BibTeX{{%
    \normalfont B\kern-0.5em{\scshape i\kern-0.25em b}\kern-0.8em\TeX}}}

\newcommand{\easylist}{EasyList}
\newcommand{\ABP}{AdBlock Plus}
\newcommand{\percentageUnusedRules}{$90.16\%$}
\newcommand{\percentageUsedRules}{$9.84\%$}
\newcommand{\numUsedRules}{$4,038$}
\newcommand{\medianLifetimeRules}{$3.8$}
\newcommand{\avgNumRulesAddedExp}{$29.8$}
\newcommand{\numRulesAddedExp}{$2,202$}
\newcommand{\numDaysCollect}{$74$}

\newcommand{\gainStrategyBlocking}{$62.5\%$}
\newcommand{\point}[1]{\par\smallskip\noindent\textbf{#1.}}

\usepackage{graphicx}
\usepackage{booktabs} 
\usepackage{makecell}

\begin{document}

\title{Who Filters the Filters: Understanding the Growth, Usefulness and Efficiency of Crowdsourced Ad Blocking}

\author{Peter Snyder}
\affiliation{
  \institution{Brave Software}
  \country{USA}
}
\email{pes@brave.com}

\author{Antoine Vastel}
\affiliation{
  \institution{University of Lille / INRIA}
  \country{France}
}
\email{antoine.vastel@univ-lille.fr}

\author{Benjamin Livshits}
\affiliation{
  \institution{Brave Software / Imperial College London}
  \country{United Kingdom}
}
\email{ben@brave.com}

\begin{abstract} 
  Ad and tracking blocking extensions are popular tools for
  improving web performance, privacy and aesthetics. Content blocking extensions
  generally rely on filter lists to decide whether a web request is associated
  with tracking or advertising, and so should be blocked.  Millions of web users
  rely on filter lists to protect their privacy and improve their browsing
  experience.

  Despite their importance, the growth and health of filter lists are poorly
  understood. Filter lists are maintained by a small number of contributors who
  use undocumented heuristics and intuitions to determine what rules should be
  included.  Lists quickly accumulate rules, and rules are rarely removed. As a
  result, users' browsing experiences are degraded as the number of stale, dead
  or otherwise not useful rules increasingly dwarf the number of useful rules,
  with no attenuating benefit.  An accumulation of ``dead weight'' rules also
  makes it difficult to apply filter lists on resource-limited mobile devices. 

  This paper improves the understanding of crowdsourced filter lists by studying
  EasyList, the most popular filter list. We measure how EasyList affects web
  browsing by applying EasyList to a sample of 10,000 websites.  We find that
  \percentageUnusedRules{} of the resource blocking rules in EasyList provide no
  benefit to users in common browsing scenarios.  We use our measurements of
  rule application rates to taxonomies ways advertisers evade
  EasyList rules.  Finally, we propose optimizations for popular ad-blocking
  tools that (i) allow EasyList to be applied on performance constrained mobile
  devices and (ii) improve desktop performance by \gainStrategyBlocking{}, while
  preserving over 99\% of blocking coverage.  We expect these optimizations to
  be most useful for users in non-English locals, who rely on supplemental
  filter lists for effective blocking and protections.
\end{abstract}

\maketitle

\section{Introduction}
\label{sec:introduction}
As the web has become more popular as a platform for information and application delivery, users have looked for ways to improve the privacy and performance of their browsing.
Such efforts include popup blockers, \texttt{hosts.txt} files that blackhole suspect domains, and privacy-preserving proxies (like Privoxy~\footnote{http://www.privoxy.org/}) that filter unwanted content.
Currently, the most popular filtering tools are ad-blocking browser extensions, which determine whether to fetch a web resource based on its URL.
The most popular ad-blocking extensions are Adblock Plus~\footnote{https://adblockplus.org/}, uBlock Origin~\footnote{https://github.com/gorhill/uBlock} and Ghostery~\footnote{https://www.ghostery.com}, all of which use filter lists to block unwanted web resources.

Filter lists play a large and growing role in making the web pleasant and useful. Studies have estimated that filter lists save users between~13 and~34\% of network data, decreasing the time and resources needed to load websites~\cite{garimella2017ad,pujol2015annoyed}.
Others, such as Merzdovnik et al.~\cite{Merzdovnik2017} and Gervais et al.~\cite{Gervais2017}, have shown that filter lists are important for protecting users' privacy and security online.
Users rely on these tools to protect themselves from coin mining attacks, drive-by-downloads~\cite{li2012knowing, zarras2014dark} and click-jacking attacks, among many others.

Though filter lists are important to the modern web, their construction is largely ad hoc and unstructured. The most popular filter lists---EasyList, EasyPrivacy, and Fanboy's Annoyance List---are maintained by either a small number of privacy activists, or crowdsourced over a large number of the same. The success of these lists is clear and demonstrated by their popularity.  Intuitively, more contributors adding more filter rules to these lists provide better coverage to users.

However, the dramatic growth of filter lists carries a downside in the form of requiring ever greater resources for enforcement. Currently, the size and trajectory of this cost are not well understood.  We find that new rules are added to popular filter lists~1.7 times more often than old rules are removed.
This suggests the possibility that lists accumulate ``dead'' rules over time, either as advertisers adapt to avoid being blocked, or site popularity shifts and new sites come to users' attention.
As a result, the cost of enforcing such lists grows over time, while the usefulness of the lists may be constant or negatively trending.
Understanding the trajectories of both the costs and benefits of these crowdsourced lists is therefore important to maintain their usefulness to web privacy, security and efficiency.

This work improves the understanding of the efficiency and trajectory of crowdsourced filter lists through an empirical, longitudinal study.  Our methodology allows us to identify which rules are useful, and which are ``dead weight'' in common browsing scenarios.   We also demonstrate two practical applications of these findings: first in optimally shrinking filter lists so that they can be deployed on resource-constrained mobile devices, and second, with a novel method for applying filter list rules on desktops, which performs \gainStrategyBlocking{} faster than the current, most popular filtering tool, while providing nearly identical protections to users.

\subsection{Research questions}
For two months, we applied every day an up-to-date version of EasyList to 10,000
websites, comprising both the 5K most popular sites on the web and a sampling of
the less-popular tail. We aimed to answer the following research questions:

\begin{enumerate}
    \item What is the growth rate of EasyList, measured by the number of rules?
    \item What is the change in the number of active rules (and rule ``matches'') in EasyList?
    \item Does the utility of a rule decrease over time?
    \item What proportion of rules are useful in common browsing scenarios?
    \item Do websites try to stealthily bypass new ad-blocking rules, and if so, how?
    \item What is the performance cost of "stale" filter rules to users of popular ad-blocking tools?
\end{enumerate}

\subsection{Contributions}
In answering these questions, we make the following primary contributions:

\begin{enumerate}
\item \textbf{EasyList over time:} we present a 9-year historical analysis of EasyList to understand the lifetime, insertion and deletion patterns of new rules in the list.
\item \textbf{EasyList applied to the web:} we present an analysis of the usefulness of each rule in EasyList by applying EasyList to 10,000 websites every day for over two months.
\item \textbf{Advertiser reactions}: we document how frequently advertisers change URLs to evade EasyList rules in our dataset and provide a taxonomy of evasion strategies.
\item \textbf{Faster blocking strategies}: we propose optimizations, based on the above findings, to make applying EasyList performant on mobile devices, and \gainStrategyBlocking{} faster on desktop environments, while maintaining over~99\% of coverage.
\end{enumerate}

\subsection{Paper organization}
The remainder of this paper is organized as follows.
Sections~\ref{sec:background} and~\ref{sec:related} provides a brief background about tracking and ad-blockers as well as a discussion of the related work.
Section~\ref{sec:easylistpast} presents a 9-year analysis of EasyList's evolution.
Section~\ref{sec:easylistpresent} presents how EasyList rules are applied on websites.
Section~\ref{sec:applicability} studies how our findings can improve ad-blocking applications on iOS and proposes two new blocking strategies to process requests faster.
Finally, in Section~\ref{sec:threats} we present the limitations,
and we conclude the paper in Section~\ref{sec:conclusion}.
\section{Background}
\label{sec:background}

\subsection{Online tracking}
Tracking is the act of third parties viewing, or being able to learn about, a
users' first-party interactions. Prior
work~\cite{Lerner2016,Englehardt2016,Yu2016} has shown that the number of
third-party resources included in typical websites has been increasing for a
long time. Websites include these resources for many reasons, including
monetizing their website with advertising scripts, analyzing the behavior of
their users using analytics services such as Google analytics or Hotjar, and
increasing their audience with social widgets such as the Facebook share button
or Twitter retweet button.

While third-party resources may benefit the site operator, they often work
against the interest of web users.  Third-party resources can harm users' online
privacy, both accidentally and intentionally.  This is particularly true
regarding tracking scripts.  Advertisers use such tracking tools as part of
behavioral advertising strategies to collect as much information as possible
about the kinds of pages users visit, user locations, and other highly
identifying characteristics.

\subsection{Defenses against tracking}
Web users, privacy activists, and researchers have responded to tracking and
advertising concerns by developing ad and tracker blocking tools.  Most
popularly these take the form of browser extensions, such as Ghostery or Privacy
badger.\footnote{https://www.eff.org/fr/node/99095} These tools share the goal
of blocking web resources that are not useful to users, but differ in the type
of resources they target.  Some block advertising, others block trackers,
malware or phishing.  These tools are popular and growing in
adoption~\cite{Mathur2018}. A report by Mozilla stated that in September 2018,
four out of the 10 most popular browser extensions on Firefox were either
ad-blockers or tracker blockers.  Adblock Plus, the most popular of all browser
extension, was used by 9\% of all Firefox
users~\footnote{https://data.firefox.com/dashboard/usage-behavior}.

Ad and tracking blockers operate at different parts of the web stack.
\begin{itemize}
  \item DNS blocking relies on a \texttt{hosts} files containing addresses of domains or sub-domains to block, or similar information from DNS. This approach can block requests with domain or sub-domain granularity, but cannot block specific URLs. Examples of domain-blocking tools include Peter Lowe's list~\footnote{http://pgl.yoyo.org/adservers/}, MVPS hosts~\footnote{http://winhelp2002.mvps.org/hosts.htm} or the Pi-Hole project~\footnote{https://pi-hole.net/}.
  \item Privacy proxies protect users by standing between the client and the rest of the internet, and filtering out undesirable content before it reaches the client.  Privoxy is a popular example of an intercepting proxy.
  \item Web browsers can attempt to prevent tracking, either through browser extensions or as part of the browser directly. These tools examine network requests and page renderings, and use a variety of strategies to identify unwanted resources.  Some tools, such as Privacy Badger use a learning-based approach, while most others use filter lists like EasyList~\footnote{https://easylist.to/pages/about.html} or EasyPrivacy~\footnote{https://easylist.to/easylist/easyprivacy.txt}.
\end{itemize}

\subsection{EasyList}
\label{sec:background:easylist}
EasyList, the most popular filter list for blocking advertising, was created
in~2005 by Rick Petnel. It has been maintained on
Github~\footnote{https://github.com/easylist/easylist} since November~2009.
EasyList is primarily used in ad-blocking extensions such as Adblock Plus,
uBlock origin and Adblock, and has been integrated into privacy oriented web
browsers\footnote{https://brave.com/}.  Tools also exist to convert EasyList
formats to other privacy tools, like
Privoxy~\footnote{https://projects.zubr.me/wiki/adblock2privoxy},

EasyList consists of tens-of-thousands of rules describing web resources that
should be blocked or hidden during display. The format also includes syntax for
describing exceptions to more general rules.  EasyList uses a syntax similar to
regular expressions, allowing authors to generalize on patterns used in URLs.

EasyList provides two categories of benefit to users.
First, EasyList describes URLs that should be blocked, or never fetched, in the browser.
Blocking resources at the network layer provides both performance benefits
(e.g. reduced network and processing costs) and privacy improvements (e.g.
reduction in number of parties communicated with or removal of fingerprinting
script code). Second, EasyList describes page elements that should be hidden at
rendering time.  These rules are useful when blocking at the network layer is not
possible.

Element hiding rules can improve the user experience by hiding unwanted page
contents, but cannot provide the performance and privacy improvements that
network layer blocking provides.

There are three types of rules in EasyList:

\begin{enumerate}
  \item \textbf{Network rules}, that identify URLs of requests that should be
    blocked.
  \item \textbf{Element rules}, that indicate HTML elements that
    should be hidden.
  \item \textbf{Exception rules}, that contradict network rules by explicitly
    specifying URLs that should not be blocked, even though
    they match a network rule.
\end{enumerate}

\begin{figure}[tb]
  \centering
  \includegraphics[width=\linewidth]{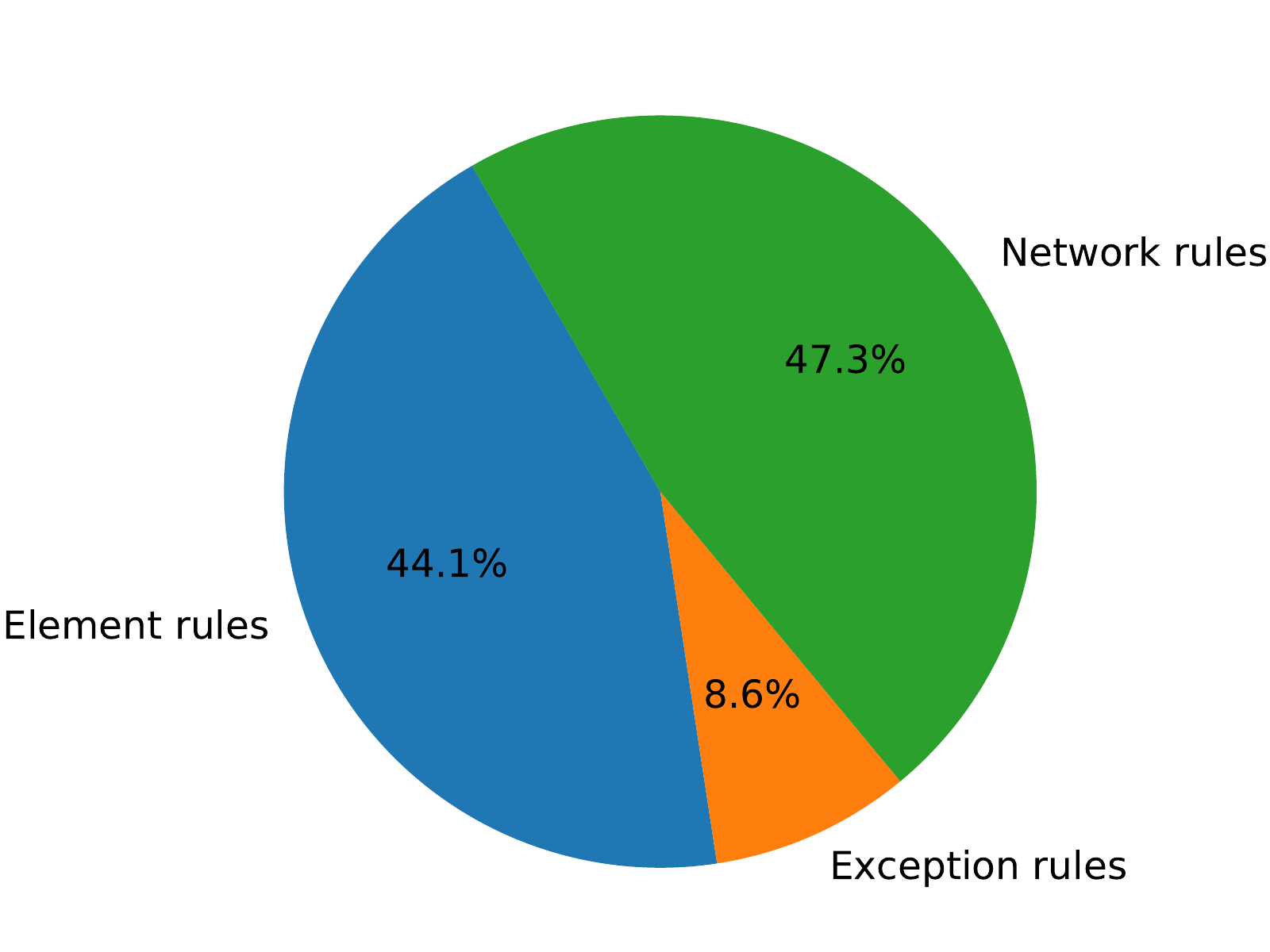}
  \caption{Distribution of rules by type in EasyList.}
  \label{graph:constitution-easylist}
\end{figure}

Figure~\ref{graph:constitution-easylist} shows the constitution of EasyList in
February 2019. Of the 71,217 rules making up EasyList, 33,703 (47.3\%) were
network rules, 6,125 (8.6\%) were exception rules, and 31,389 (44.1\%) were
element rules.

\section{Composition of EasyList Over Time}
\label{sec:easylistpast}
This section measures how EasyList has evolved over its 9-year history, through
an analysis of project's public commit history.  The section proceeds by first
detailing our measurement methodology, along with our findings that EasyList has
grown to comprise nearly 70,000 rules and that it is primarily maintained by a
very small number of people. The section concludes by showing that most rules
stay more than \medianLifetimeRules{} years in EasyList before they are removed,
suggesting a huge accumulation of unused rules.

\subsection{Methodology}
EasyList is maintained in a public repository on GitHub.  We use
\textbf{GitPython}~\footnote{https://gitpython.readthedocs.io/en/stable/index.html},
a popular Python library, to measure commit patterns and authors in the EasyList
repository over the project's 10-year history.

\subsubsection{Measurement frequency}
For every commit in the EasyList repository, we record the author and the type
and number of rules modified in the commit.

First we grouped commits by day.  We then checkout each day commits and measure
which rules have been added and removed since the previous day.  We use this
per-day batching technique to avoid artifacts introduced by \texttt{git diff},
which we found causes over-estimations of the number of rules changed between
commits.

\subsubsection{Accounting for changes in repository structure}
The structure of the EasyList repository has changed several times over the
project's history.  At different times, the list has been maintained in one file
or several files.  The repository has also included other distinct-but-related
projects, such as EasyPrivacy.  We use the following heuristics to attribute
rules in the repository to EasyList.

When the repository consists of a single \texttt{easylist.txt} file, we check
to see if the file either contains references (e.g.  URLs or file paths) to
lists hosted elsewhere, or contains only filter rules.  When the
\texttt{easylist.txt} file contains references to other lists, we treat EasyList
as the union of all rules in all referenced external lists.  When
\texttt{easylist.txt} contains only filter rules, we treat EasyList as the
content of the \texttt{easylist.txt} file.

When the repository consists of anything other than a single
\texttt{easylist.txt} file, we consider EasyList to be the content of all the
files matching the following regular expression \texttt{'easylist\_*.txt'} and
that are located in the main directory or in a directory called
\texttt{easylist}.

\subsection{Results}

\subsubsection{Rules inserted and removed}
\begin{figure}[tb]
  \centering
  \includegraphics[width=\linewidth]{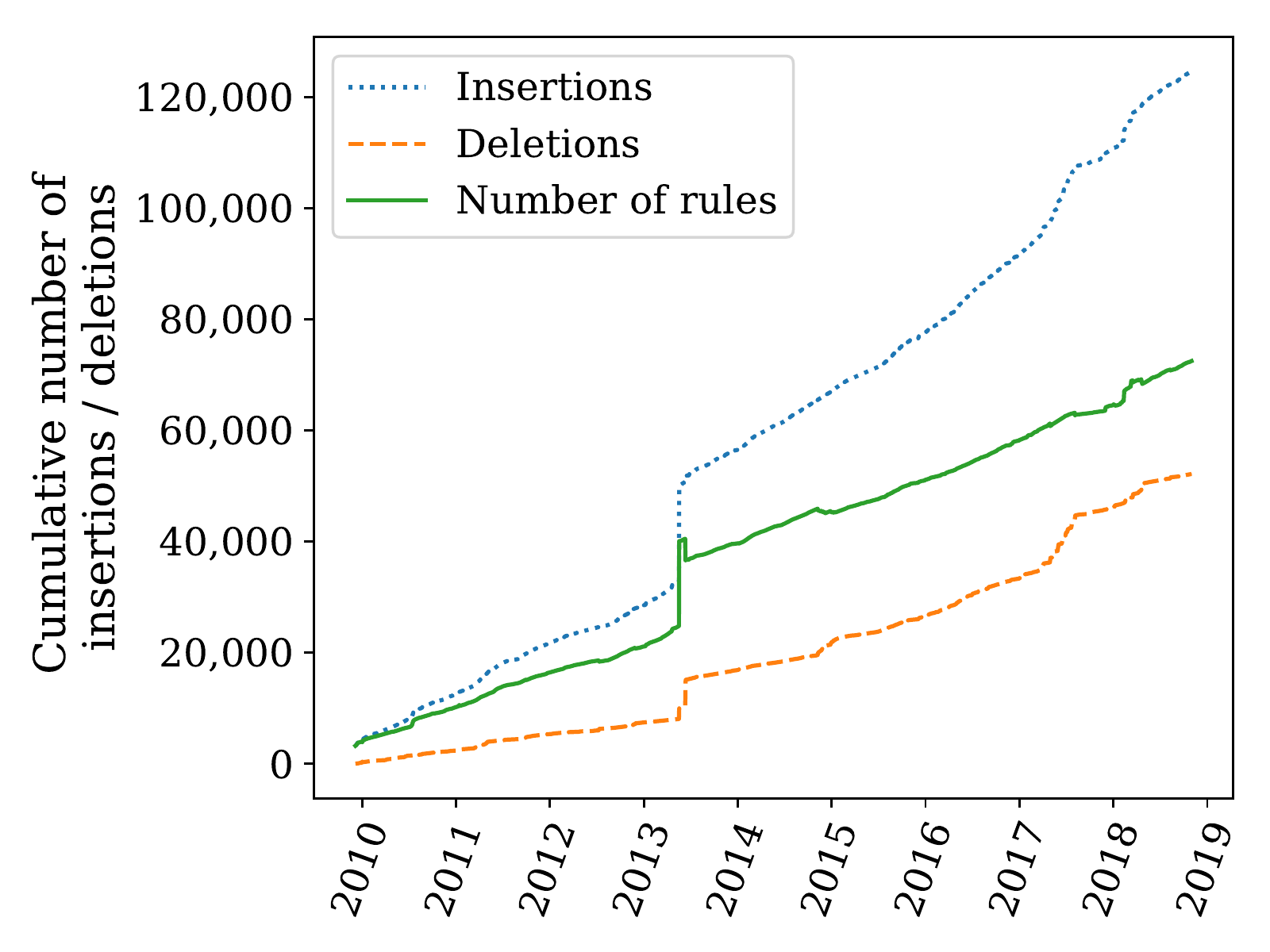}
  \caption{Evolution of the number of rules in EasyList. Over a 10-year period, EasyList grew by more than 70,000 rules.}
  \label{graph:evolution-easylist}
\end{figure}

Figure~\ref{graph:evolution-easylist} presents the change in the size of
EasyList over time.  It shows the cumulative number of rules inserted, removed,
and present in the list over nine years.  Rules are added to the list faster
than they are removed, causing EasyList to grow larger over time.  Over a 9-year
period, 124,615 rules were inserted and 52,146 removed, resulting in an increase
of 72,469 rules.  EasyList's growth is mostly linear.  One exception is the
sharp change in 2013, when ``Fanboy's list'', another popular filter list, was
merged into EasyList.

\subsubsection{Modification frequency}
We analyzed the distribution of the time between two commits in EasyList and
observed that EasyList is frequently modified, with a median time between
commits of 1.12 hours, and a mean time of 20.0 hours.

\subsubsection{EasyList contributors}
Contributors add rules to EasyList in two ways.  First, potential contributors
propose changes in the EasyList forum.\footnote{https://forums.lanik.us/}
Second, contributors can post issues on the EasyList Github repository.

Though more than 6,000 members are registered on the EasyList forum, we find
that only a small number of individuals decide what is added to the project. The
five most active contributors are responsible for 72,149 of the 93,858 (76.87\%)
commits and changes.  65.3\% of contributors made less than 100 commits.

\subsubsection{Lifetime of EasyList rules}
\begin{figure}[tb]
  \centering
  \includegraphics[width=\linewidth]{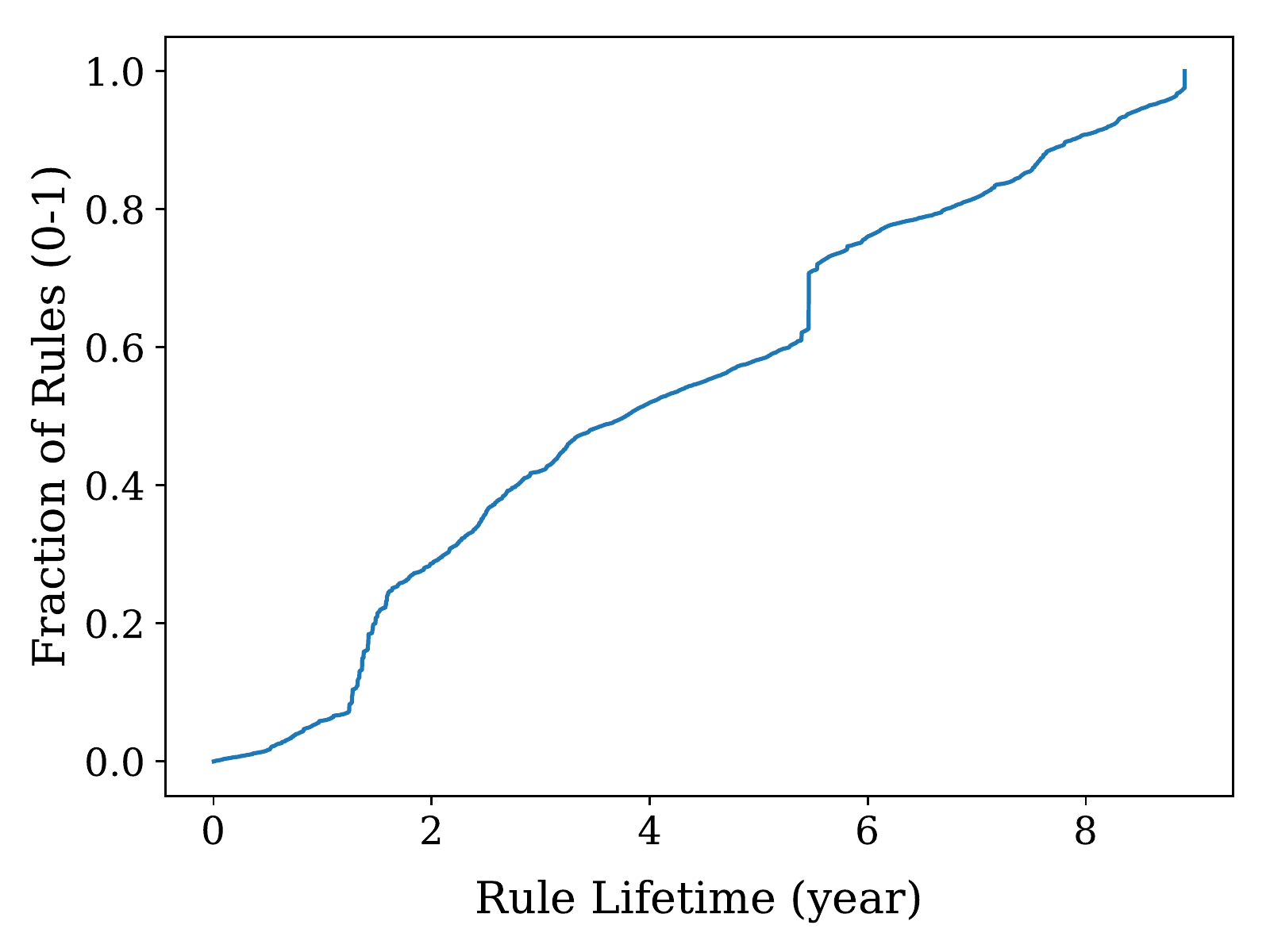}
  \caption{Cumulative distribution function of the lifetime of the rules in
    EasyList. Half of the rules stay more than \medianLifetimeRules{} years in
    EasyList.}
  \label{graph:cdf-lifetime-easylist}
\end{figure}

Figure~\ref{graph:cdf-lifetime-easylist} presents the distribution of the
lifetime of rules in EasyList.  The figure considers only rules that were
removed during the project's history.  Put differently, the figure shows how
much time passed between when a rule was added to EasyList, and when it was
removed, for the subset of rules that have been removed. We observe that 50\% of
the rules stayed more than \medianLifetimeRules{} years (45.5 months) in
EasyList before being removed.

\section{EasyList applied to the Web}
\label{sec:easylistpresent}

This section quantifies which EasyList rules are triggered in common browsing
patterns.  We conducted this measurement by applying EasyList to 10,000 websites
every day for two months, and recording which rules were triggered, and how
often.

This section first presents the methodology of a longitudinal, online study of a
large number of websites.  We then present the results of this measurement.  We
find that over 90\% of rules in EasyList are never used.  We also show that on
average, \avgNumRulesAddedExp{} rules were added to the list every day, but that
these new rules tend to be less used than rules that have been in EasyList for a
long time.

The section concludes by categorizing and counting the ways advertisers react to
new EasyList rules.  We detect more than 2,000 situations where URLs are changed
to evade rules and present a taxonomy of observed evasion strategies.

\subsection{Omitting element rules}
The results presented in this section describe how often, and under what
conditions, network and exception rules apply to the web.  However, as discussed
in Section~\ref{sec:background:easylist}, EasyList contains \textit{three}
types of rules: network rules that block network requests, exception rules
that prevent the application of certain network rules, and element rules that
describe parts of websites that should be hidden from the user for cosmetic
reasons.

We omit element rules from our measurement for three reasons.  First, our
primary concern is to understand how the growth and changes in EasyList
affect the privacy and security of Web browsing, and element rules have
no effect on privacy or security. In all modern browsers,
hiding page elements has no affect on whether those elements are fetched
from the network and, in the case if \texttt{<iframe>} elements, rendered in
memory~\footnote{In fact, many Web applications use this quirk of hidden
iframes to create simplified, early versions of server-push communication,
an approach called ``long polling''. Similarly, tracking scripts like Google
Analytics use never-rendered, never-visible images for client-server
communication.}. In all but uncommon edge cases, relating mostly to memory
availability, hidden elements are still fetched from network, though possibly
with a lower priority.

Second, we omit element rules from the study because their application is highly
variable, and would add a not-useful amount of dimensionality to the data.
Element rules apply differently depending on how users interact with the page,
since page layouts can change in response to user interaction and timer events.
This is especially true in highly dynamic, client-side web applications, like
those written in JavaScript frameworks like
React~\footnote{https://reactjs.org/} and
Angular~\footnote{https://angularjs.org/}, since client-side page
modifications can change which element rules apply. Network rules, in contrast,
have far less (though not zero) variability over the life-cycle of a
page~\footnote{There are exceptions here, such as analytic scripts
that initiate network requests to recorder, server side, user behaviors.
However, note that
these are relativity uncommon, since the initial analytic script is generally
blocked.}.

Third, though least significantly, we omit element rules from consideration
because there are common uses of EasyList where element rules are not applied,
lessening the value of measuring this portion of the list. The most common
examples of such uses are privacy-preserving network proxies
(e.g. Privoxy~\footnote{https://www.privoxy.org/} and
SquidGuard~\footnote{http://www.squidguard.org/index.html}).

\subsection{Methodology}
\begin{table}[!t]
\centering
\setlength{\tabcolsep}{6pt}
\begin{tabular}{lr}
    \toprule
        \bf Measurement                 & \bf Counts \\
    \midrule
        \# days                      & 74 \\
       	\# domains                   & 10,000 \\
        \# non-responsive domains    & 400 \\
    \midrule
        Avg \# pages per day         & 29,776 \\
        Avg \# pages per domain per day & 3.74 \\
        Total \# pages measured      & 3,260,479 \\
    \bottomrule
\end{tabular}
\caption{Statistics of the number of domains and sites measured during online EasyList measurement.}
\label{table:online-measurement-stats}
\end{table}
This subsection discusses how we measured how EasyList effects typical web
browsing, including what sites were measured, the instrumentation used tool
take the measurements, and what information was collected.  The following
subsection describes the results of executing this methodology.

\subsubsection{Crawl description}
To understand the usefulness of rules in EasyList, we applied EasyList to 10,000
websites every day for over two months (\numDaysCollect{} days).  We selected
these 10,000 websites from two groups:

\begin{enumerate}
    \item \textbf{Popular websites:} Websites from the top 5K Alexa, a
        ranking of sites online by popularity.

    \item \textbf{Unpopular websites:} 5,000 websites randomly selected from
        the top Alexa one-million, but not present in the set of popular
        websites. (i.e. rank 5,001--1 million)
\end{enumerate}

We crawled the web using an instrumented version of Chromium to measure which
filter rules were applicable when browsing a large number of websites in an
anonymous browsing scenario.  The crawls were launched from AWS Lambda instances
located in the \texttt{us-east-1} region.

For each day of the experiment, we first visit the landing page of each URL in
the popular and unpopular sets. We then randomly selected up to three URLs
referenced in anchor tags, pointing to pages on the eTLD+1 domain, and visit
these URLs.  This resulted in between 10,000 and 40,000 pages being measured
every day.  Table \ref{table:online-measurement-stats} provides high level
statistics of these measurements.

We use the Chrome devtools
protocol~\footnote{https://chromedevtools.github.io/devtools-protocol/} to
record the following information about each network request made during page
execution:

\begin{itemize}
    \item Time of the request
    \item URL of the request
    \item URL of the domain that initiated the request
    \item Type of resource fetched (e.g. image, script, sub-documents)
    \item Hash of the response
    \item Size of the response
\end{itemize}

To avoid introducing side effects, we did not block any requests during our
measurements.  We instead first recorded all HTTP requests made when interacting
with each site. Next, we applied EasyList offline using Brave's ad-blocker
NodeJS module~\footnote{https://github.com/brave/ad-block}.  We determine if
each request would have been blocked, excepted, or allowed.  A ``blocked''
request is one that matches an EasyList network rule, indicating that the
resource should not be fetched.  An ``excepted'' request is one that matches a
blocking rule, but also matches a ``excepting'' rule, indicating that the
resource \textit{should} be fetched anyway. An ``allowed'' request matches no
EasyList rules.

\subsubsection{Description of the dataset}
Our dataset comprises every network request made during our crawl of~10,000
websites, for \numDaysCollect{} days between July 24th, 2018 and October~5th,
2018.  Among the 10,000 websites crawled daily, 400 (4.0\%) never responded
during the experiment.  We attribute this to a mix of websites becoming inactive
(common among unpopular websites~\cite{scheitle2018long}) and websites blocking
IP addresses belonging to AWS to deter crawlers.  The existence of such
AWS-including IP blacklists has been documented in other
work~\cite{invernizzi2016cloak}.  We discuss possible limitations to our study
more in Section~\ref{sec:threats}.

\subsection{Results}
\subsubsection{Proportion of EasyList rules used}
We consider a rule as used during a crawl if the rule matched at least one
network request made during the crawl.  We measure the proportion of network and
exception rules used during our crawls.  As noted above, we measure network and
exception rules, but not element (e.g. cosmetic) rules because network rules
impact performance and privacy.

We find that the vast majority of network and exception rules are never used.
Only \percentageUsedRules{} (\numUsedRules) of rules were used even once during
our measurements.

An even smaller number of network and exception rules are \textit{frequently}
used.  On average, only 5.14\% of EasyList network and exception rules were used
at least once per day (\textbf{Research Question 4}). We also observed that the
number of active rules is stable over time (\textbf{Research Question 2}).

\begin{figure}[tb]
  \centering
  \includegraphics[width=\linewidth]{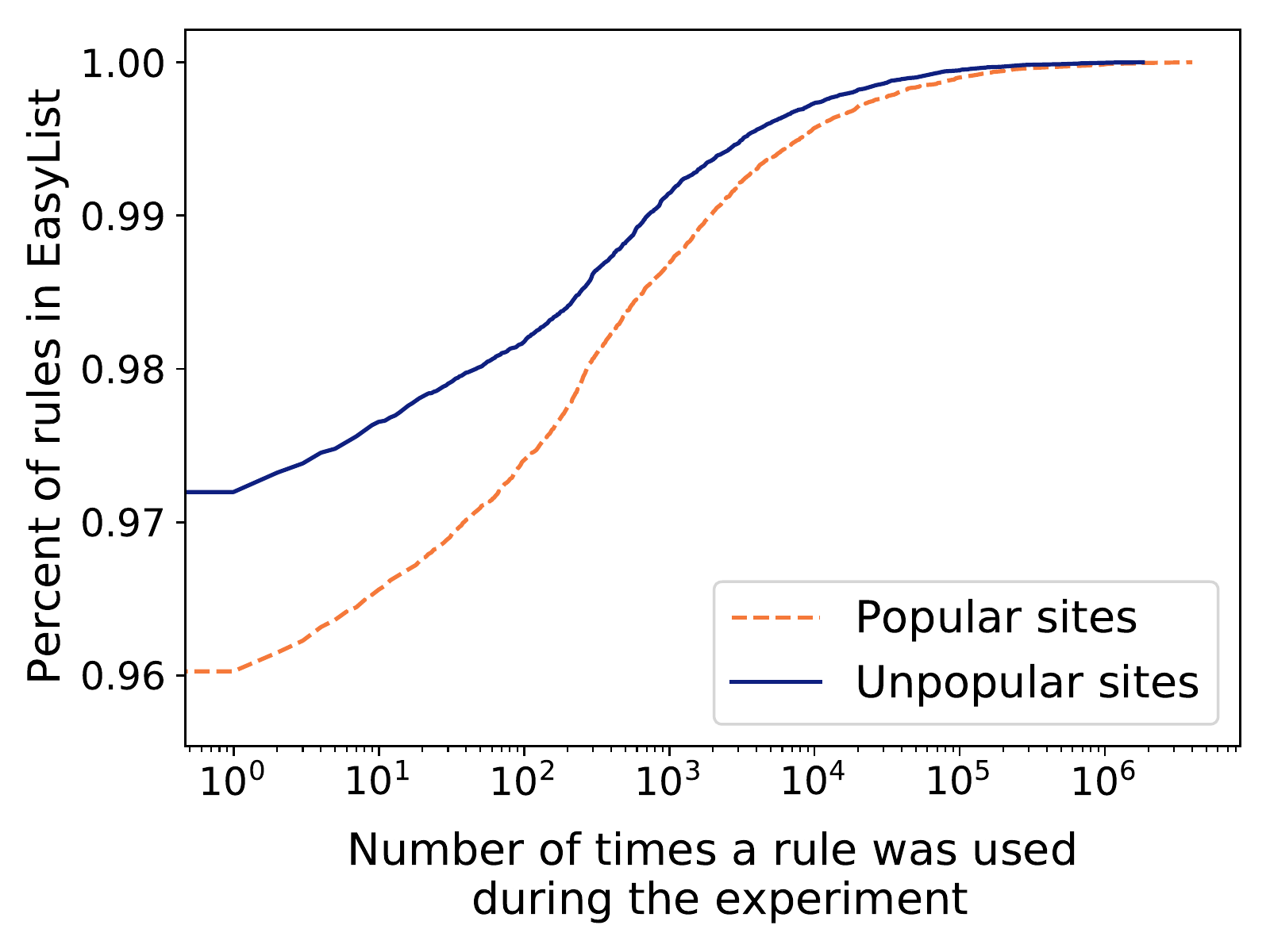}
  \caption{Distribution of the number of times rules were used during the
  \numDaysCollect{} days of the experiment, on popular (Alexa rank 1--5k) and
  unpopular (Alexa rank 5,001--1m).}
  \label{graph:combined-rule-cdf}
\end{figure}

Figure~\ref{graph:combined-rule-cdf} shows the cumulative distribution
function of how often filter rules were used during the \numDaysCollect{} days
of the experiment. The distribution is skewed; the majority of rules are either
not used (\percentageUnusedRules), or were used between 1 and 100
times (4.45\%).  Only 3.56\% of the rules were used between 100 and 1,000 times,
and~1.83\% more than 1000 times.

\subsubsection{Usefulness of EasyList additions}
During the experiment, \numRulesAddedExp{} network and exception rules were
added to EasyList, an average of \avgNumRulesAddedExp{} new rules per day
(\textbf{Research Question 1}). We refer to rules added to EasyList during our
measurement campaign as \textit{new}; we call rules \textit{old} if they existed
in EasyList at the start of the measurement period.

The vast majority of rules, new and old, were not used during our measurements.
Of the \numRulesAddedExp{} rules added during the study period, 208 (9.45\%)
were used at least once.  Those measurements are roughly similar for old rules
(\percentageUsedRules).  However, when considering only rules that were used at
least once, we found that new rules were used nearly \textit{an order of
magnitude less} than old rules.  This suggests a declining marginal usefulness
per rule as EasyList accumulates more rules, possibly because the most
troublesome resources are already blocked. If a new rule was used during the
study, it was used an average of 0.65 times per day.  Old rules were applied
much more frequently, 6.14 times a day on average.

\subsubsection{Impact of the age of rules}
We also measure whether the age of a rule impacts its use.  We find that the age
of a rule significantly impacts the number of times rules are used.  We present
these findings in two ways: graphically and statistically.

\begin{figure}[tb]
  \centering
  \includegraphics[width=\linewidth]{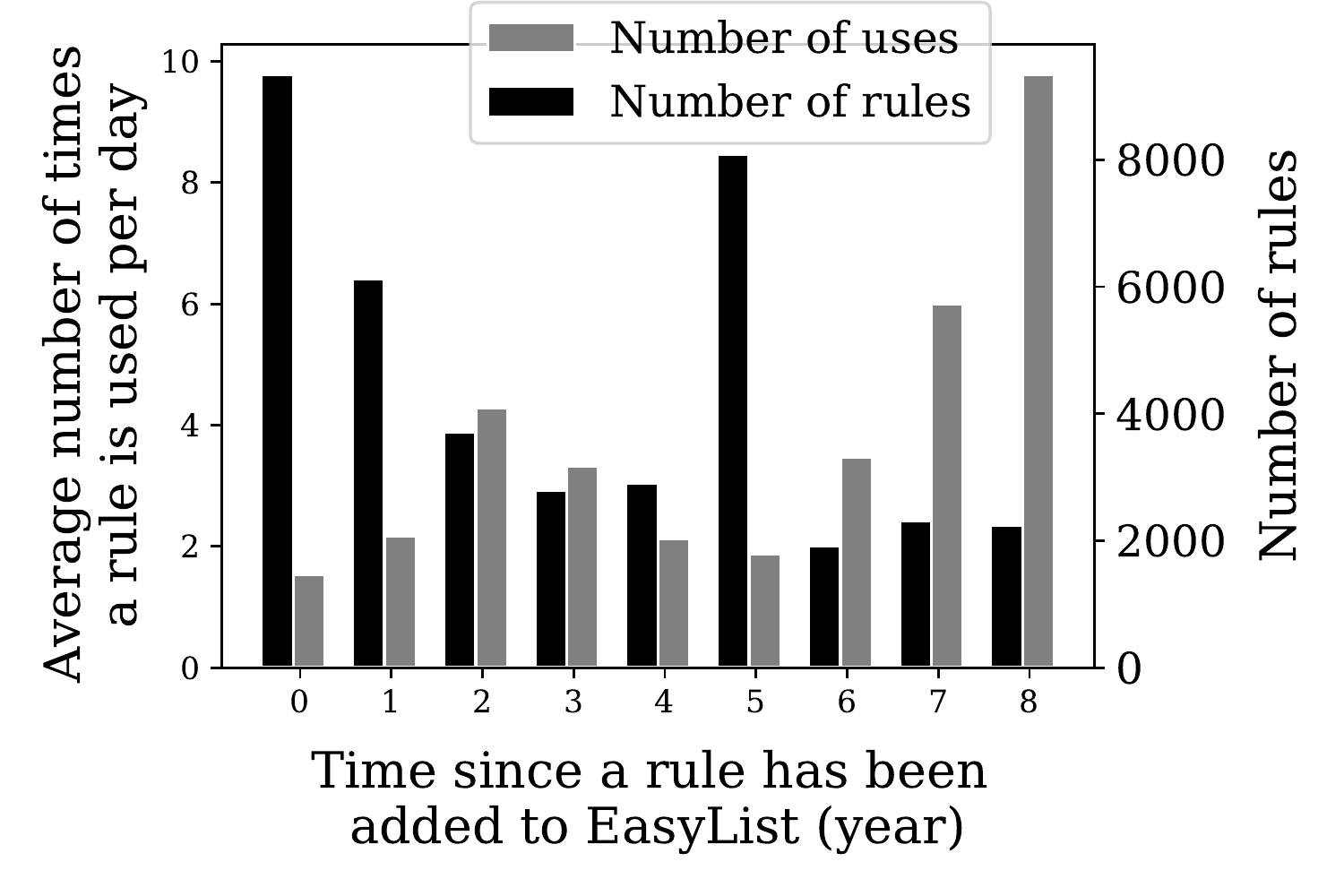}
  \caption{The black bars represent the distribution of the age of the rules present in EasyList. The grey bars represent the average time a rule is used per day against the time it has been added to \easylist{}.}
  \label{graph:avg-use-age}
\end{figure}

Figure~\ref{graph:avg-use-age} presents the distribution of the age of the rules
present in EasyList, as well as the average number of uses depending on the age
of a rule.  The distribution of the age of the rules (black bars) shows that
there are rules from all ages in EasyList.  The important number of 5-year-old
rules is explained by the merge of the "Fanboy's list", another popular filter
list, with EasyList in 2013.  If we observe the average number of times rules
are used in a day (grey bars), we see that the most useful rules are old.  This
is caused by generic rules blocking URLs that contained keywords such as "ads"
or popular domains such as \url{doubleclick.net}. For example, the rule
\texttt{.com/ads/\$image,object,subdocument} was added to EasyList in
September 2010 and was triggered $748,330$ times during the experiment.  We did
not observe a linear relationship between the average use of a
rule and its age.

Besides the graphical analysis, we also conduct statistic tests to determine
whether the age of a rule impact its use. We use the Kolmogorov-Smirnov test to
compare the distribution of the number of times rules are used depending on the
duration they have been present in EasyList. For each year $y_i$ between 1 and
8, we compare the distribution of the number of times rules that have been
present $y_i$ years in EasyList have been used with:

\begin{enumerate}
    \item the distribution of the number of times rules that have been present
        less than one year,
    \item the distribution of the number of times rules that have been present
        between $y_{i-1}$ and $y_i$ years.
\end{enumerate}

For all the tests, we obtain p-values less than $10^-5$, which indicates that
there are significant differences in the way rules are used depending on their
age (\textbf{Research Question 3}).

\subsection{Advertiser Reactions}
\label{subsec:reaction}
EasyList helps users avoid online advertising.  Advertisers and websites that
rely on advertising for their income do not want their content to be blocked and
may try to circumvent rules in EasyList.  There are several ways an advertiser
may do this.  One option is to try to detect the use of an adblocking
tool~\cite{Iqbal2017}.  Alternatively, the advertiser may manipulate the URLs
that serve their content, to prevent matching filter rules.  In this section, we
measure how often, and in which ways, advertisers responded to EasyList rules.
We do not observe a statistically significant reaction by advertisers \textit{in
general} but we note common patterns in avoidance strategies among a subset of
advertisers (\textbf{Research Question 5}).

We measured advertiser reactions to EasyList rules through the following
intuition: if a resource changed URL more frequently after being blocked by
EasyList than before, it suggests an advertiser trying to evade EasyList.
Similarly, if the number of times a resource was blocked spiked after the
EasyList addition, and then reverted to its pre-rule block-rate, that would also
suggest advertiser evasion.

We used this intuition through the following steps.  First, we considered only
rules that were added during the measurement period, and which remained in
EasyList for at least 14 days.  Second, we identified resources that were
blocked by these new rules and looked to see if the same resource (as determined
by the content's hash), was served by different URLs during the study.  Third,
we filtered out resources that were less than 50KB, to avoid resources that were
trivially identical, like common logos and tracking pixels.  Fourth, we measured
whether the number of URLs a resource was served from changed significantly
before and after being blocked by EasyList.

\begin{figure}[tb]
  \centering
  \includegraphics[width=\linewidth]{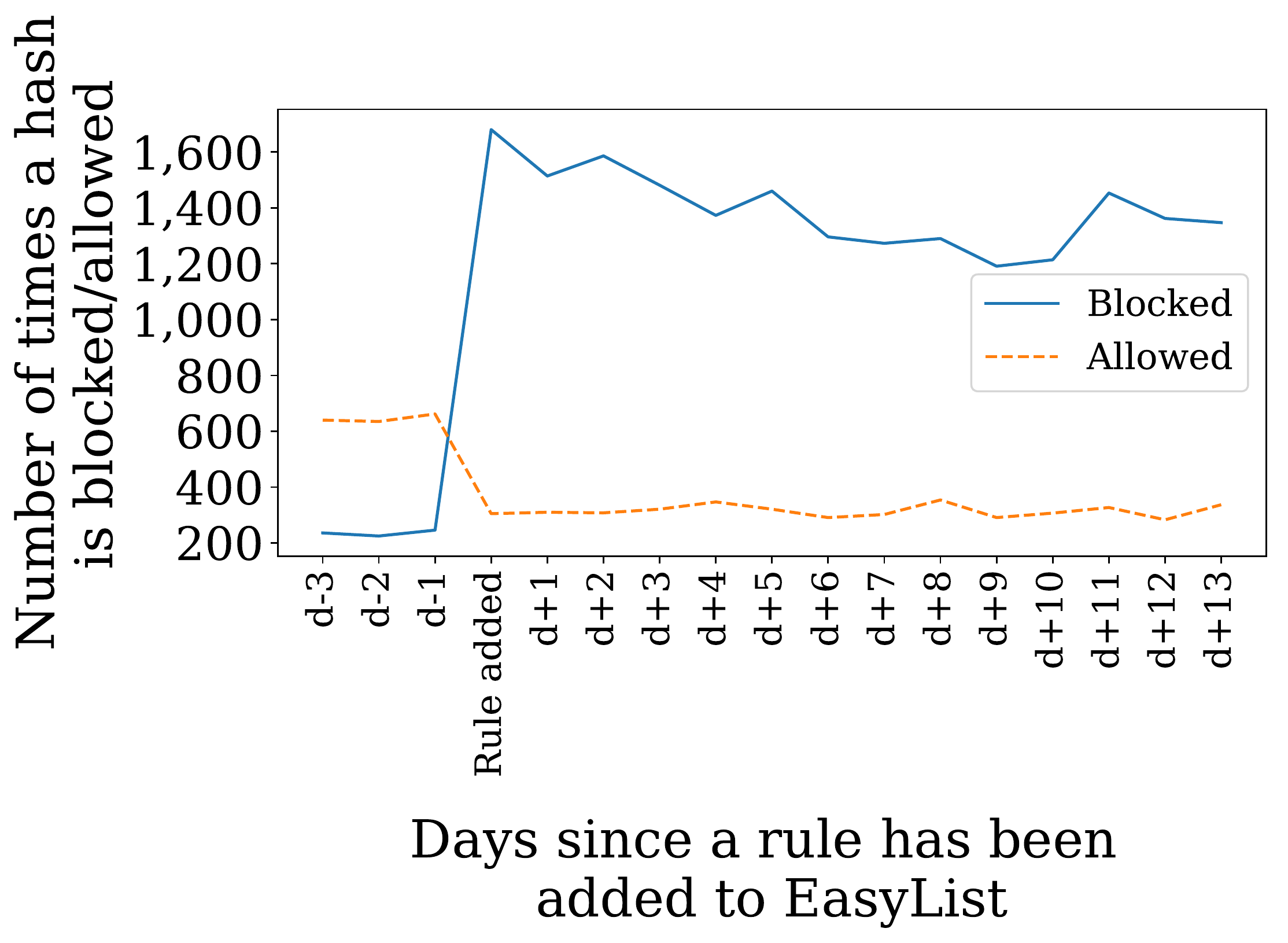}
  \caption{Number of times resources are blocked or allowed after a rule is added to \easylist{}}
  \label{graph:global-reaction-websites}
\end{figure}

Figure~\ref{graph:global-reaction-websites} presents the block and allow rates
for resources affected by new rules.  We did not find any population-wide trends
of advertisers modifying URLs to avoid EasyList.  If advertisers were, in
general, successfully evading EasyList, we would observe a decrease in blocking
over time. Block rates did not though, in general, revert to pre-rule levels
over time.

\subsection{Evasion Strategies}
\begin{figure}[tb]
  \centering
  \includegraphics[width=\linewidth]{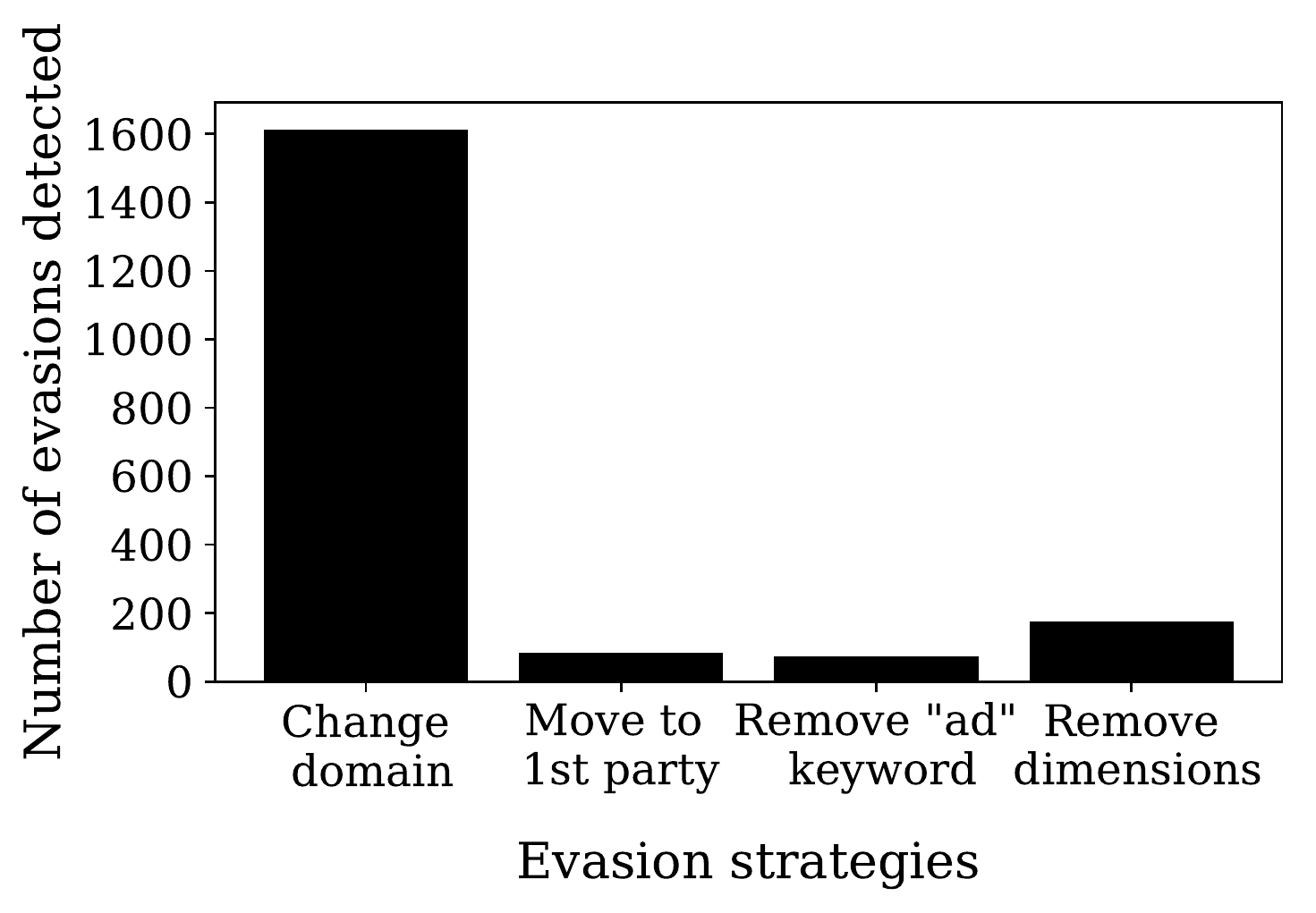}
  \caption{Number of times each strategy has been used during the experiment.}
  \label{graph:count-evasion-strategies}
\end{figure}

In this subsection, we present a partial taxonomy of the strategies used by
advertisers to avoid EasyList rules.
Figure~\ref{graph:count-evasion-strategies} presents the number of times each
evasion strategy was observed.

\subsubsection{Changing Domains}
Many advertisers changed the domains their resources were served from, either by
subtly modifying the domain names or by completely changing them through domain
generation algorithm style techniques.  This happened 1,612 times and does not
take into account resources that were moved to the first party.  For example,
the URL \url{https://c.betrad.com/geo/ba.js?r170201} was blocked by the rule
\texttt{||betrad.com\^{}\$third-party}.  The resource was moved to a new domain,
\url{c.evidon.com}, to avoid being blocked.

In total, resources were moved from 100 distinct domains to 157 distinct
domains, representing a total of 195 distinct combinations of domains.  The two
most frequent transitions are resources moved from
\url{pagead2.googlesyndication.com} to \url{google.com} and from
\url{cs03.etcodes.com} to \url{cs03.et-cod.com}.  The former occurred 499 times
and the latter 185 times.  We observe that the advertiser changed
the domain that served the resources so that it looks the same when observed by
a human but that does not match the filter anymore.

\subsubsection{Moving Resources to the First Party}
Advertisers avoided EasyList rules by moving resources from third-party domains
to the first party.  It happened 84 times, among which 23 times resources were
moved to another sub-domain of the first party.  For example, we observed the
domain \url{cnn.com} including resources from \url{ssl.cdn.turner.com}, which
was blocked by the rule \texttt{||turner.com\^{}*/ads/}.  We then observed the
same resource being served from \url{cdn.cnn.com} directly, which prevented the
resource from matching the \texttt{||turner.com} domain in the filter rule.

\subsubsection{Removing Ad Keywords from URLs}
Keywords such as `ads' or `ad' trigger filter rules.  We observed $73$ URLs
where these keywords were simply removed.  For example, the URL
\url{https://etherscan.io/images/ad/ubex-20.png} was blocked by the rule
\texttt{/images/ad/*}.  To bypass the filter rule, the URL was changed to
\url{https://etherscan.io/images/gen/ubex-20.png}.

\subsubsection{Removing Image Dimensions from URLs}
Some URLs contain parameters to specify the dimension of the ad banners.  These
parameters also trigger filter rules.  Advertisers can evade these rules by
removing matching parameters from URLs.  We observed $176$ Evasions of this
kind.  For example,
\url{https://s0.2mdn.net/dfp/.../lotto_kumulacja\_160x600\_009/images/lotto_swoosh.png}
was blocked by the rule \texttt{\_160x600\_}.  We observed an advertiser
removing the dimension parameter from their URLs to avoid being blocked.

\section{Applications}
\label{sec:applicability}

In this section, we present two practical applications of the previous sections'
findings: first, an optimized, reduced EasyList
on resource constrained iOS devices, and second, a novel EasyList-based
filtering strategy targeting desktop extensions, that provides nearly all of the
blocking benefits of full EasyList, but with significantly improved performance.

\subsection{Improving Content Blocking on iOS}
We first present how content blocking in iOS differs from content blocking in
other platforms.  Then, we run a benchmark that measures how the size of a
filter list impacts the time to launch a content-blocking application on iOS,
and how our findings could help to decrease this time.

\subsubsection{Overview of Content Blocking on iOS}
iOS and Safari use a different strategy for content blocking than other
browsers~\footnote{https://developer.apple.com/documentation/safariservices/creating\_a\_content\_blocker}.
On most platforms, ad-blocking tools receive information about each request,
such as the request URL, the expected resource type, etc.
The extension can then apply whatever logic is desired to
determine whether the request should be blocked. In most content blocking
systems, this results in a large number of regular expressions (or similar text
patterns) applied to the URL, along with some optimization to limit the number
of rules that need to be considered.

iOS and Safari (along with Google's proposed Manifest v3
changes\footnote{https://docs.google.com/document/d/1nPu6Wy4LWR66EFLeYInl3NzzhHzc-qnk4w4PX-0XMw8/edit})
use a different approach, where extensions declare a static set URL patterns
that should be blocked but \textit{do not execute the rule application logic}.
This protects the user from malicious extensions (since extensions
cannot modify requests with malicious code), at the cost of requiring all rules
be expressed in a format \textit{that is less expressive than the
EasyList format}.  The result is that EasyList is generally expanded from
EasyList's compact rule format to a larger number of iOS-compatible rules.

\point{Limitations} There are two relevant limitations in iOS's blocking approach.
First, iOS enforces a limit of 50K rules.  This limit is not in Apple's
documentation, but the main ad-blocking applications report it
\footnote{https://help.getadblock.com/support/solutions/articles/6000099239-what-is-safari-content-blocking-,
https://www.ublock.org/blog/introducing-ublock-safari-12/}, and we observe the
same during testing. Since Easylist alone contains 40K network rules, little
room is left for either other popular lists (e.g. EasyPrivacy) or region specific
EasyList supplements (e.g. EasyList China). iOS's restrictions thus limit the
amount of protection users can deploy.

This limit on the number of rules is particularly harmful to non-English speaking
users, who often need to rely on supplemental, region or language specific
filter lists, that are applied in addition to EasyList.  As EasyList itself
is large enough to consume the iOS limit on filter rules, non-English
users cannot use EasyList with their regional list, resulting in reduced
protections~\cite{sjosten2020regionalfilters}.

Second, iOS compiles filter rules into a binary format each time rules are
updated.  As we show in the benchmark we conduct, users may have to wait for 14
seconds or more when a list composed of 40K rules is compiled.
This is particularly unacceptable when launching an app for
the first time when tricks like background compilation cannot be used to hide
the cost from users.

\subsubsection{Benchmark}\hfill
\point{Approach} We use the findings of this work to decrease rule compilation
cost (and increase the ability of users to include other lists of rules) on iOS
devices by only compiling filter rules that are likely to be useful, instead of
the full set of 40k rules.  We show that reducing EasyList to only its useful
rules provides a dramatically improved initial launch experience for users, and
gives users more flexibility to apply additional filter lists. The primary
beneficiaries of this optimization are non-English speakers, and those on
reduced capability mobile devices.

\point{Evaluation Methodology} We first measure the costs of compiling different
sizes of filter lists on different popular iOS devices.  We generate lists that
contain between 1,000 and 40,000 rules randomly selected from the set of network
and exception rules in EasyList.  For each of the lists, we use a fork of the
``ab2cb'' library~\footnote{https://github.com/brave/ab2cb} to convert the rules
from the Adblock Plus format to the iOS JSON format.  Then, for each device and
each list, we compile each iOS filter list 5 times and report the average
compilation time.

\begin{figure}[tb]
  \centering
  \includegraphics[width=\linewidth]{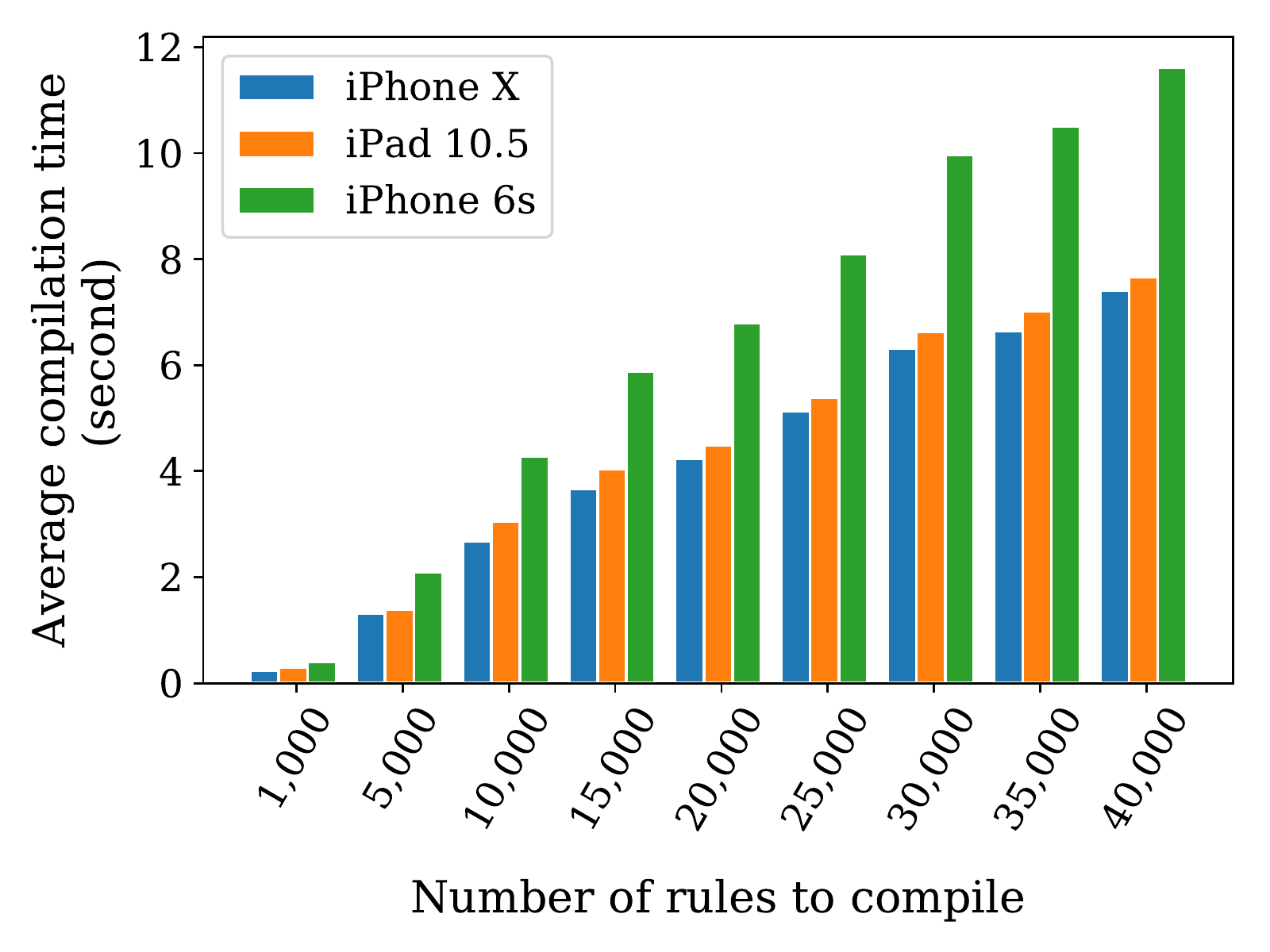}
  \caption{Average time to compile a filter list depending on device and size of the list.}
  \label{graph:bench-ios}
\end{figure}

\point{Results}
Figure~\ref{graph:bench-ios} shows the average compilation time
for each device and each selected list size.  The compilation times grows
linearly with respect to the number of rules.  While on average it takes~$0.24$
second to compile a list composed of 1,000 rules on an iPhone~X, it grows
to~$7.4$ seconds for a list composed of~40,000 rules.

Device kind also impacts compilation time. Compiling 40k rules on
an iPhone 6s takes on average 11.62s, 4.2 seconds longer than on an iPhone X.

Thus, keeping only active rules has two main benefits in the case of ad-blockers
running on iOS and Safari.  First, it allows users to enjoy the benefits of
EasyList while remaining well under the platform's 50K rule-limit.  Second,
it dramatically decreases the required compilation time for the first time the
application is launched.

\subsection{Improving ad-blocking performance}
\label{sec:applicability:adblock-perf}

We also propose an optimized strategy for applying EasyList that provides nearly
all of the benefits of traditional ad-blockers while improving filter list
application speed.  We describe our technique in three steps. First, we describe
how current tools use EasyList for blocking (using AdBlock Plus, the most
popular of such
tools~\footnote{https://data.firefox.com/dashboard/usage-behavior}, as
a representative example).  Second, we present a ``straw''
blocking strategy that considers only frequently used filter rules.  Third, we
propose a novel hybrid strategy that achieves nearly the accuracy of existing
techniques with the performance improvements of the ``straw'' strategy.

Our hybrid approach achieves over~99\% of the coverage of the current most
popular EasyList tool, but performs \gainStrategyBlocking{} faster. Because of
the nature of the optimizations in this hybrid strategy, we expect it could be
applied to other EasyList consuming tools to achieve similar performance
improvements. This hybrid approach achieves this performance improvement
at the cost of some user privacy, since resources blocked by infrequently used
EasyList rules would still be loaded once. We note that this privacy
``cost'' is only paid once, while the performance improvement is ongoing, and so
might be an appealing trade off to even privacy-sensitive users.

\point{Strategy One: Synchronous Full EasyList}
Most EasyList tools decide whether a network request should be blocked as
follows:

\begin{enumerate}
  \item Use hardcoded heuristics, such as not blocking top-level
    documents or requests coming from non-web protocols.  If any of these
    heuristics match, allow the request.

  \item Check the requested URL against the small number ``exception'' rules
    in EasyList.  If any ``exception'' rules match, allow the request.

  \item See if the requested URL matches any of the ``network'' rules in
    EasyList.  If any ``network'' rule matches, block the request.

  \item Otherwise, allow the request.
\end{enumerate}

We note two performance impacting aspects of this strategy.  First, it performs
a large number of unnecessary computation, since every ``exception'' and
``network'' rule in EasyList is applied to outgoing request, even though the
vast majority (over \percentageUnusedRules) are very unlikely to be useful
(again, based on the measurements described in
Section~\ref{sec:easylistpresent}). Second, this wasteful computation adds delay
to a time-sensitive part of the system. These filter checks are conducted
synchronously, blocking all outgoing network requests until all EasyList rules
are considered.

\point{Strategy Two: Synchronous Reduced List} Next, we describe a straw-man
strategy, that improves performance by only considering the
\percentageUsedRules{} of rules expected to be useful.  This strategy is
otherwise identical to strategy one and differs only in the number of EasyList
rules considered. Instead of applying all of EasyList's 38,710 ``network'' and
``exception'' rules, this strategy only evaluates the \numUsedRules{} rules
observed during the online measurements discussed in
Section~\ref{sec:easylistpresent}.  The expected trade-off here is performance
for coverage since resources that match rarely used filters will be allowed.

\begin{figure}[tb]
  \centering
  \includegraphics[width=0.7\linewidth]{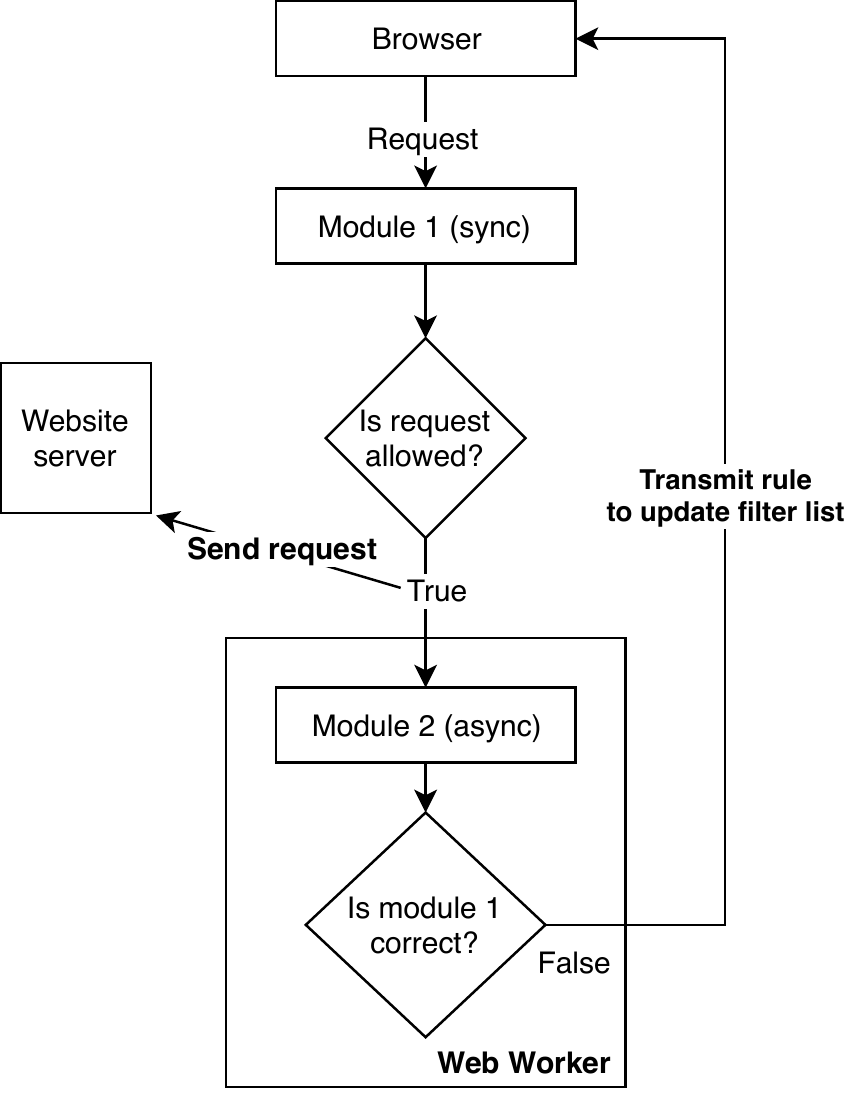}
  \caption{Overview of the proposed hybrid strategy.}
  \label{graph:strategy2}
\end{figure}

\point{Strategy Three: Synchronous Reduced List, Asynchronous Complementary
List} Finally, we present our proposed blocking strategy, a hybrid strategy that
achieves nearly the coverage that full EasyList provides while achieving the
performance improvements of the reduced list. Figure~\ref{graph:strategy2}
outlines this hybrid strategy.

This strategy uses two steps:

\begin{enumerate}
  \item A synchronous, request-time matcher, that operates before each request
    is issued, but with a reduced version of EasyList.
  \item An asynchronous background matcher, that applies the uncommon tail of
    EasyList, but only when a request has been allowed by the previous step.
\end{enumerate}

The first step is identical to strategy two.  Every outgoing network request is
intercepted and blocked until the frequently-used subset of EasyList rules is
considered. The step's goal is to minimize how long network requests are
blocked, by minimizing the amount of synchronous work.  The result is that
benign network requests complete more quickly than current blocking tools.

The second step applies the remaining, long-tail of EasyList rules, but at a
less performance-sensitive moment. If a network request is allowed by the first
step, the request is issued, but the browser continues checking the now-issued
URL against the rest of EasyList.  This continued checking is done
asynchronously so that it has minimal effect on the load time of the page.

If the asynchronous checker finds any rules that match the URL of the now-issued
request, that rule is added to the set of rules applied by the synchronous
matcher so that it will be quickly blocked in the future.

The result of this hybrid strategy is that commonly blocked requests are blocked
quicker (because the synchronous blocking step is considering a smaller rule
set), benign requests complete faster (again, because of the reduced rule list
used in the synchronous blocker), and rare-but-undesirable URLs are adjusted to
(because the asynchronous matcher moves matching filter rules into the
synchronously-applied set).

\subsubsection{Evaluation Methodology}

We evaluated the performance of each strategy by implementing them in \ABP{} for
Chrome~\footnote{https://github.com/adblockplus/adblockpluschrome}.
In all strategies, we instrumented \ABP{} to measure the time needed
to evaluate each outgoing network request against EasyList (or the relevant
subset of EasyList).  For the hybrid approach, we also added timing measurements
to the asynchronous step.

We evaluated each of the three blocking strategies against the same selection of
popular and unpopular websites discussed in Section~\ref{sec:easylistpresent}.
We conduct the crawl on an AWS T2 medium instance with 2 virtual CPUs and 4GB of
RAM.  Since Chromium does not support extensions in headless
mode,~\footnote{https://bugs.chromium.org/p/chromium/issues/detail?id=706008} we
use stock Chromium rendered with XVFB, and automated the system with the
Puppeteer library~\footnote{https://github.com/GoogleChrome/puppeteer}. All
experiments were conducted with caching disabled.

For each strategy we visited each of the 10K websites in our
sample. We allowed each website five seconds to respond and then allowed each
website to execute for two seconds. The extension measures the time taken by
\ABP{} (modified or stock) to decide whether to block each network request on
each page.

\subsubsection{Performance of Blocking Strategies}
\begin{table*}[t]
\small
\centering
\renewcommand{\arraystretch}{1.2}
\setlength{\tabcolsep}{2.1pt}
\begin{tabular}{llrrrrrrrr}
\toprule
&\thead{Strategy} & \thead{Num rules\\ start} & \thead{Num rules\\ end} & \thead{Median eval time \\per request (ms)} & \thead{90th time \\per request (ms)} & \thead{Num requests \\blocked} & \thead{Num requests \\exceptioned} & \thead{Num 3rd parties\\ contacted} & \thead{Num total \\requests} \\
\midrule
\textbf{(1)} & Easylist sync & 39,132 & 39,132 & 0.30 & 0.50	& 30,149 &	11,905 & 322,604	& 748,237 \\
\textbf{(2)} & Reduced list sync          & 3,259            & 3,259          & 0.10                            & 0.30                       & 30,611                & 2,508                     & 340,301                      & 774,473             \\
\textbf{(3)} & Hybrid combined    & 39,132           & 39,132       & 0.30                       & 0.60                       & 31,584                & 14,444                    & 338,841                      & 770,729            \\
\textbf{(3.1)} & Hybrid sync  & 3,259            & 3,445  & 0.20                         & 0.30                       & 31,446                & 14,396                    & -                      & -             \\
\textbf{(3.2)} & Hybrid async & 35,873           & 35,687    & 0.20                         & 0.30                       & 138                  & 48                       & -                           & -           \\
\bottomrule
\end{tabular}
\caption{Performance and coverage comparison for three EasyList application strategies.}
\label{table:blocking-strategies}
\end{table*}

Table~\ref{table:blocking-strategies} presents the results of applying the above
evaluation methodology against each of the three blocking strategies.

The first row presents measurements for the stock \ABP{} implementation, which
uses a synchronous blocking strategy for all of EasyList.  Unsurprisingly, this
strategy takes the longest time to determine whether to block a network request.
We note that this time is spent blocking each network request, which greatly
impacts page load time.

The second row shows the performance of the second strategy; reducing EasyList
to its most frequently used rules, and applying that reduced list synchronously.
The result is faster performance, at the cost of an increased false positive
false negative rate. The number of network requests blocked
\textit{goes up} because of ``exception'' pruned from EasyList.  Privacy
is also harmed, as nearly~18,000 more third-parties are contacted during the
evaluation, a result of some ``network'' rules missing in the reduced EasyList.

The remaining rows present the evaluation of the hybrid
strategy. Rows four and five describe the synchronous and
asynchronous modules of the hybrid strategy separately, while row three presents
the combined effect.  The most significant results of our evaluation are the
following.

First, the synchronous module takes~0.21 ms on average to process a request.
Perceived blocking time is reduced (compared to stock \ABP) by
\gainStrategyBlocking{} (\textbf{Research Question 6}).  Second, the hybrid
strategy provides blocking coverage nearly identical to stock \ABP~($>99\%$),
with only~$138$ false negatives on~$10,000$ websites visited. The~$48$
``exception'' rule errors do not impact the user since the rules that would have
been excepted were not added to EasyList.  Third, the evaluation shows the
adaptive benefit of the hybrid model.  The hybrid approach initially
applied~3,259 rules synchronously, but after the~10,000 site evaluation, 186
rules from the uncommon async set were added to the synchronous set.

Second, we note the asynchronous portion of the hybrid
approach applies its~35K rules faster than the reduced-list synchronous
approach, which considers only~3,259 rules.  This surprising observation is
due to the synchronous portion of the hybrid approach doing some work
(e.g. what kind of resource is being fetched) that can be reused in the
asynchronous step.

Overall, we find that the hybrid approach is a successful, performant strategy.
The hybrid strategy considers only the subset of EasyList that is likely to be
useful in the performance critical path, and defers evaluating the
less-likely-to-be-useful rules to a less critical decision point.  The hybrid
approach achieves these improvements with a minimal effect on blocking accuracy,
and at a small (though not zero) privacy cost, on the order of one non-blocked
resource per uncommonly used filter list rule.

Finally, we note that there are potential further performance improvements
that might be achieved by pushing the hybrid approach further, and starting
each user with an empty set of filter rules.  This would increase
the privacy cost of the hybrid approach, since a larger number of
would-be-blocked resources would be fetched, but would result in an even
smaller, further optimized set of rules that tightly matched each users'
browsing patterns.

\section{Limitations and Discussion}
\label{sec:threats}

\subsection{Web site selection generalizability}
The findings in this study depend on having a sample of the web that generalizes
to the types of websites users visit and spend time on.  We treat the Alexa 5K,
along with a sampling of less popular websites, as representative of typical
browsing patterns.  While we expect this set to be a good representative of the
web as a whole (largely because the highly skewed distribution of website
popularity means the most popular sites represent the majority of most user's
browsing time), we note it here as a limitation, and that extending this
work's measurements beyond the Alexa 5k would be valuable future work.

Additionally, this work considers each site's landing page, and up to three
sub-pages linked to from the site's landing page, as representative of the
site's content overall.  If this is not the case, and deeply nested pages
have different kinds and amounts of resources than higher-level pages, it would
reduce the generalizability of this work's findings.  We note this as an
area for future work.

\subsection{Web region and language generalizability}
This work applies EasyList globally popular sites, as determined by the Alexa
global rankings.  This choice was made because EasyList itself targets English
and "global" sites.  Other lists target others languages and regions on the web.
Some of these lists are maintained by the EasyList project
\footnote{https://easylist.to/pages/other-supplementary-filter-lists-and-easylist-variants.html},
others lists are created by other filtering tools \footnote{e.g.
https://kb.adguard.com/en/general/adguard-ad-filters} or crowdsource
efforts \footnote{e.g. https://filterlists.com/}. It would be interesting
future work to understand how EasyList performs on other regions of the web
(as compared to English and ``global'' sites), and how EasyList's performance
compares to region-and-language-specific lists.

\subsection{Automated measurement generalizability}
All of our results were generated from automated crawls, which also may have
affected how generalizable our results are.  It is possible that different kinds
of resources are fetched and so different parts of EasyList are used, when users
interact with websites in particular ways, such as logging in or using web-app
like functionality.  How generalizable automated crawl results are to the
browsing experiences of real users is a frequently acknowledged issue in
measurement studies (e.g. \cite{snyder2017most}), and one we hope the community
can address with future work.

Additionally, all crawling done in this work was carried out from well known AWS
IP addresses.  This means that the results may be affected by the kinds of
anti-crawling techniques sometimes deployed against Amazon IP addresses.  This,
in turn, could have affected the number and distribution of ads observed during
measurement.  While this is a common limitation of this kind of web-scale
measurement, we note it as another limitation.

\subsection{Relationship to filter list evasion}
Many websites prefer for their included resources not be blocked by filter
lists, for reasons ranging from monetization to anti-fraud efforts. Some
sites and advertisers attempt to evade filter lists (and other blocking tools).
Some of these techniques are presented in Section \ref{subsec:reaction}, and
others have been detailed in other research and discussed in Section
\ref{subsec:list-maintenance}.

While, if effective, these evasion efforts would would reduce the usefulness of
filter-list based blocking, evasion efforts are unlikely to be effective in the
common case. First, advertisers and trackers are constrained in their ability to
evade filter lists because frequently changing URLs would break existing sites
that have hard coded references to well known URLs.  Second, frequently changing
URLs imposes a non-zero cost on advertisers and trackers by making caching
difficult, and so increasing serving costs.  Finally, though trackers may
consider evading filter lists with more expensive techniques (e.g.  domain
generation algorithms, resource inlining, etc), many may be hesitant to do so
because, in the long term, more sophisticated efforts will likely be defeated
too, since the client has the ultimate ability to choose what
content to fetch and render, for reasons described by Storey et.
al.~\cite{Storey2017}.

\subsection{Varying resource blocking importance}
Finally, our results consider every blocking action as equally useful.  In our
measurements, a rule that blocks ten resources is implicitly ten times more
useful than a rule that only blocks one request.  It is possible, though, that a
less frequently used rule may be more beneficial to the user than a frequently
used rule if the infrequently blocked rule is blocking a very malicious or
performance harming resource.  While we expect the most severe, security harming
resources are most commonly dealt with through other blocking tools, such
as SafeBrowsing~\footnote{https://safebrowsing.google.com/}, we
acknowledge this limitation, but treat the more general question of
``how beneficial is it to the user to block a given resource'' beyond the scope
of this work.

\section{Related work}
\label{sec:related}

\subsection{Online tracking}
Recent studies document a growth of third-party tracking on the
web~\cite{Lerner2016,Englehardt2016}.  Yu et al.~\cite{Yu2016} found an increase
in analytics services and social widgets.  Englehardt et
al~\cite{Englehardt2016} showed tracking code on more than 10\% of the websites
of the top Alexa 1M.  Libert et al~\cite{Libert2015} showed that 180k pages of
the top 1M Alexa websites had cookies spawned by the DoubleClick domain, a
behavioral advertising company.

\subsection{Defenses against tracking}
The NoScript extension~\footnote{https://noscript.net/} enables to prevent
JavaScript execution.  While this approach blocks trackers, it also breaks
websites that use JavaScript for legitimate purposes.  This trade-off between
privacy and usability is important.  Yu et al.~\cite{Yu2016} find that privacy
tools that break legitimate websites may lead to users deactivating such tools,
harming user privacy.  The most popular kind of tracking protection is browser
extensions such as Ghostery, Disconnect or uBlock origin, as well as browsers
such as Brave or Safari that enable to block third-party requests.

A variety of strategies have been proposed for identifying unwanted web
resources. Privacy Badger uses a learning-based approach.  Iqbal et
al.~\cite{Iqbal2018} also proposed a machine learning approach that considers
features extracted from HTML elements, HTTP requests, and JavaScript to
determine if a request should be blocked.  Storey et al.~\cite{Storey2017}
propose a visual recognition approach targeting legally mandated advertising
identifiers.  Yu et al~\cite{Yu2016} proposed a crowdsourced approach where
users collectively identify data elements that could be used to uniquely
identify users. The majority of anti-tracking and ad-blocking tools rely on
filter lists.

Different studies~\cite{pujol2015annoyed} show that tracker
blockers and ad-blockers are popular among the general population.  Malloy et
al~\cite{Malloy2016} showed that depending on the country, between 16\% and 37\%
of the Internet users had an ad-blocker installed.  Mathur et
al~\cite{Mathur2018} found that most users of anti-tracking tools use the tools
to avoid advertising.

\subsection{Effectiveness of anti-tracking tools}
Gervais et al.~\cite{Gervais2017} quantified the privacy provided by the main
ad-blockers.  They show that on average, using an ad-blocker with the default
configuration reduce the number of third parties loaded by 40\%.  Merzdovnik et
al.~\cite{Merzdovnik2017} showed that rule-based approaches can outperform
Privacy Badger's learning-based approach.  They also show that extensions that
rely on community-based lists are less effective than extensions based on
proprietary lists such as Disconnect or Ghostery when used with the correct
settings.  Their study demonstrates that besides blocking trackers, most of
these extensions have a negligible CPU overhead.  In some cases, it even leads
to a decrease in the overall CPU usage of the browser.

\subsection{Maintaining filter lists}
\label{subsec:list-maintenance}
In order to keep up with new domains creating and domains changing their
behavior, it is crucial to maintain filter lists.  Because it is a cumbersome
task and it needs to be done carefully not to break websites, Gugelmann et
al.~\cite{Gugelmann2015} proposed an automated approach that relies on a set of
web traffic features to identify privacy invasive services and thus help
developers maintaining filter lists.

Alrizah et al.~\cite{alrizah2019errors} studied the related problem of
how filter lists maintainers detect and address blocking errors, and how
advertisers attempt circumvent filter lists.  They find that popular
lists have non-trivial false positive and false negative rates, and that
these errors are exploited by attackers (i.e. advertisers).

Other researchers have also documented strategies advertisers use to evade
blocking.  Wang et al~\cite{Wang2016} found advertisers randomizing
HTML identifiers and structure.  Facebook has applied this technique
too~\footnote{\url{https://newsroom.fb.com/news/2016/08/a-new-way-to-control-the-ads-you-see-on-facebook-and-an-update-on-ad-blocking/}}.  Adversity has also been discussed by recent
studies on anti-ad-blockers~\cite{Iqbal2017, Zhu2018, nithyanandKJVFP16}, i.e.
scripts whose purpose is to detect and block ad-blockers to deliver advertising
to more users.  Iqbal et al.~\cite{Iqbal2017} conducted a retrospective
measurement study of anti ad-block filter lists using the Wayback machine, and
found that 8.7\% of popular sites have at one time used anti-adblocking
scripts.

\section{Conclusion}\label{sec:conclusion}
This paper studies EasyList, the most popular filter list used for blocking advertising and tracking related content on the web.  We find that the vast majority of rules in EasyList are rarely, if ever, used in common browsing scenarios.  We measure the number of these ``dead weight'' rules, and effect on browser performance, through comparison with alternative, data-driven EasyList application strategies.  We find that by separating the wheat from the chaff, and identifying the small subset of EasyList filter rules that provide common benefit for users, EasyList's benefits can be efficiently enjoyed on performance constrained mobile devices.  We also use these findings to propose an alternate blocking strategy on desktops that improves performance by \gainStrategyBlocking{}, while capturing over $>$ 99\% of the benefit of EasyList.

More broadly, we hope this work will inform how similar crowdsourced security and privacy tools are developed and maintained.  As previous work~\cite{katz2018toward} has identified, such lists tend to accumulate cruft as they accumulate new rules.  Over time, the benefit of such tools risks being outweighed by the amount of dead weight pulled with them.  We hope the findings in this work highlight the need for regular pruning of these lists, to keep them lean and as helpful to users as possible.
\section{Acknowledgements}
\label{sec:ack}
The authors would like to thank Yan Zhu, who suggested the core of the
optimization described in Section \ref{sec:applicability:adblock-perf},
and Ryan Brown (i.e. "Fanboy"),
a core EasyList maintainer who was helpful throughout this project.

\bibliographystyle{plain}
\bibliography{paper}

\end{document}